 \newlength\smallfigwidth
\documentclass[aps,prb,twocolumn,floatfix,showpacs,amsmath,amssymb,nofootinbib,superscriptaddress] {revtex4-2}

\usepackage{lineno}
\usepackage{amsmath}    
\usepackage{graphicx}   
\usepackage{verbatim}   
\usepackage{color}      
\usepackage{subfigure}  
\usepackage{float}
\usepackage{hyperref}   

\newcommand{\orcid}[1]{\href{https://orcid.org/#1}{\raisebox{2.5pt}{\includegraphics[width=7pt]{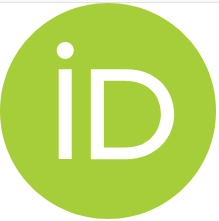}}}}

\usepackage[usenames,dvipsnames]{xcolor}

\definecolor{MS-color}{rgb}{0.4,0.8,0.3}

\renewcommand{\selectlanguage}[1]{}

\begin{document}


\title{Superconducting Spintronic Heat Engine}

\author{C. I. L. de~Araujo\orcid{0000-0003-2801-1759}}
\email{dearaujo@ufv.br}
\affiliation{NEST, Istituto Nanoscienze-CNR and Scuola Normale Superiore, I-56127 Pisa, Italy}
\affiliation{Departamento de F\`{i}sica, Laborat\'{o}rio de Spintr\^{o}nica e Nanomagnetismo, Universidade Federal de Vi\c{c}osa, Vi\c{c}osa,36570-900, Minas Gerais, Brazil}
\author{P. Virtanen\orcid{0000-0001-9957-1257}}
\email{pauli.t.virtanen@jyu.fi}
\affiliation{Department of Physics and Nanoscience Center, University of Jyv\"askyl\"a, P.O. Box 35 (YFL), FI-40014 University of Jyv\"askyl\"a, Finland}
\author{M. Spies\orcid{0000-0002-3570-3422}}
\affiliation{NEST, Istituto Nanoscienze-CNR and Scuola Normale Superiore, I-56127 Pisa, Italy}
\author{Carmen González-Orellana\orcid{0000-0003-4033-5932}}
\affiliation{Centro de F\'{i}sica de Materiales (CFM-MPC), Centro Mixto CSIC-UPV/EHU, 20018 Donostia-San Sebasti\'{a}n, Spain} 
\author{Samuel Kerschbaumer\orcid{0000-0001-6099-2265}}
\affiliation{Centro de F\'{i}sica de Materiales (CFM-MPC), Centro Mixto CSIC-UPV/EHU, 20018 Donostia-San Sebasti\'{a}n, Spain} 
\author{Maxim Ilyn\orcid{0000-0001-5832-2449}}
\affiliation{Centro de F\'{i}sica de Materiales (CFM-MPC), Centro Mixto CSIC-UPV/EHU, 20018 Donostia-San Sebasti\'{a}n, Spain} 
\author{Celia Rogero\orcid{0000-0002-2812-8853}}
\affiliation{Centro de F\'{i}sica de Materiales (CFM-MPC), Centro Mixto CSIC-UPV/EHU, 20018 Donostia-San Sebasti\'{a}n, Spain} 
\affiliation{Donostia International Physics Center (DIPC), 20018 Donostia-San Sebasti\'{a}n, Spain}
\author{T. T. Heikkilä\orcid{0000-0002-7732-691X}}
\affiliation{Department of Physics and Nanoscience Center, University of Jyv\"askyl\"a, P.O. Box 35 (YFL), FI-40014 University of Jyv\"askyl\"a, Finland}
\author{F. Giazotto\orcid{0000-0002-1571-137X}}
\email{francesco.giazotto@sns.it}
\affiliation{NEST, Istituto Nanoscienze-CNR and Scuola Normale Superiore, I-56127 Pisa, Italy}
\author{E. Strambini\orcid{0000-0003-1135-2004}}
\email{elia.strambini@cnr.it}
\affiliation{NEST, Istituto Nanoscienze-CNR and Scuola Normale Superiore, I-56127 Pisa, Italy}

\begin{abstract}

Heat engines are key devices that convert thermal energy into usable energy. 
Strong thermoelectricity, at the basis of electrical heat engines, is present in superconducting spin tunnel barriers at cryogenic temperatures where conventional semiconducting or metallic technologies cease to work.  
Here we realize a superconducting spintronic heat engine consisting of a ferromagnetic insulator/superconductor/insulator/ferromagnet tunnel junction (EuS/Al/AlO$_x$/Co). 
The efficiency of the engine is quantified for bath temperatures ranging from 25 mK up to 800 mK, and at different load resistances. Moreover, we show that the sign of the generated thermoelectric voltage can be inverted according to the parallel or anti-parallel orientation of the two ferromagnetic layers, EuS and Co. This realizes a thermoelectric spin valve controlling the sign and strength of the Seebeck coefficient, thereby implementing a thermoelectric memory cell.
We propose a theoretical model that allows describing the experimental data and predicts the engine efficiency for different device parameters.

\end{abstract}


\maketitle

\section{Introduction}
Thermoelectricity can be observed when an electron-hole-asymmetric conductor is driven by a temperature difference. 
The resulting thermovoltages and thermocurrents have been widely applied for thermometry and energy harvesting\cite{disalvo1999thermoelectric,snyder2008complex,pelegrini2014development}.
In the current technology, semiconductors have been extensively utilized due to the large Seebeck coefficient achievable in those materials, allowing for amplitudes of up to hundreds of $\mathrm{\mu V / K}$ \cite{sootsman2009new,roddaro2013giant,prete_thermoelectric_2019} and efficient energy harvesting \cite{vullers2009micropower}.
However, semiconductor materials are not ideal for thermoelectric-based applications in some important areas of research such as aerospace and cryogenic electronics. At the extremely low temperatures of space or in dilution cryostats, the carriers in semiconductors freeze out\cite{wachutka1990rigorous} and the material becomes insulating.
Approaches based on quantum dot system have been proposed and realized \cite{josefsson_quantum-dot_2018,prete_thermoelectric_2019} showing sizable thermopower down to $0.5$K but with limitations of scalability intrinsic to zero-dimensional systems.
Moreover semiconductor properties would be drastically changed by the doping due to background cosmic particles  \cite{keyes2001fundamental}.

\begin{figure*}[hbt!]
\center
\includegraphics[width=0.95\textwidth]{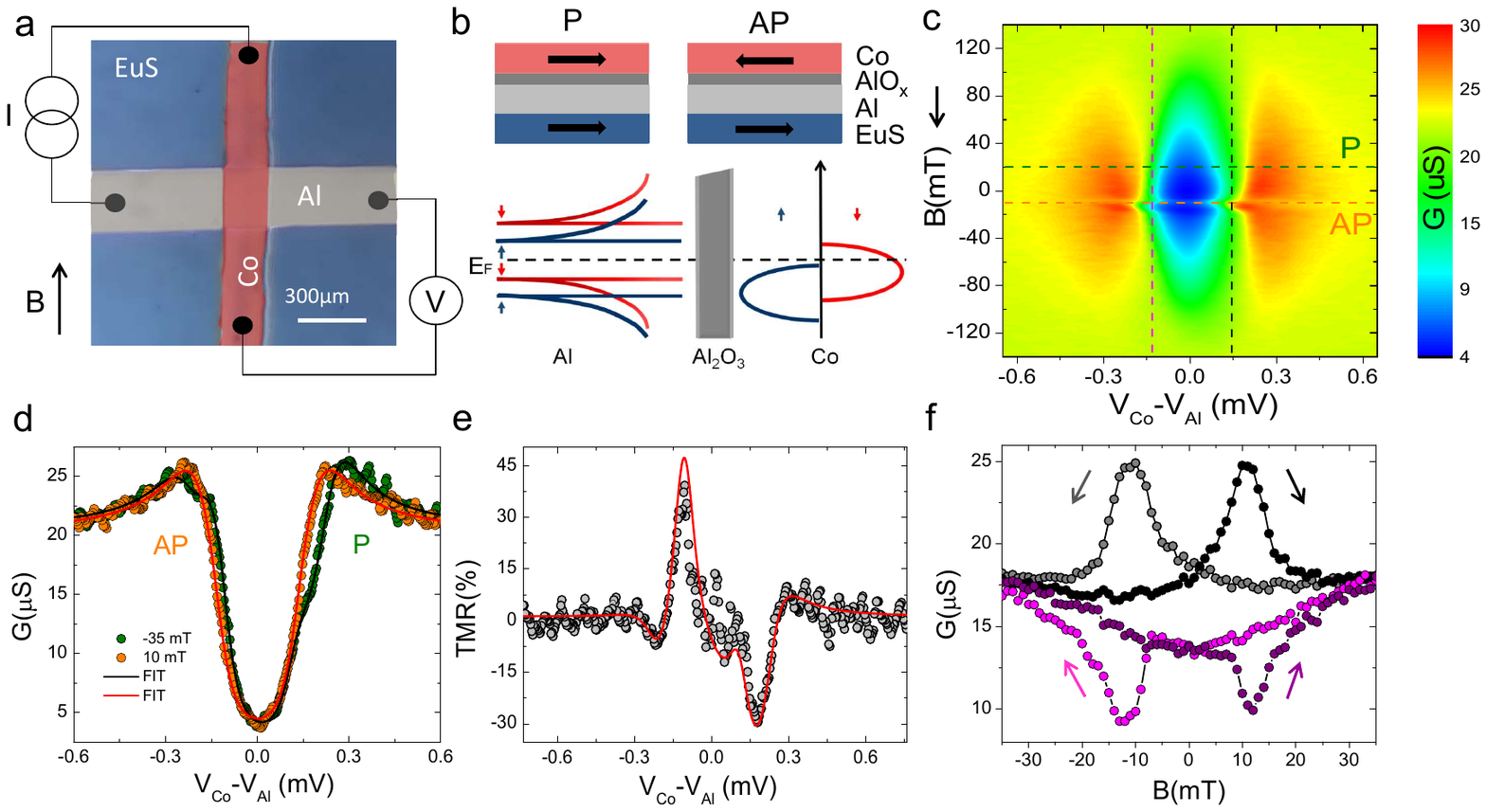}
\caption{{\bf Magneto-electric characterization of the superconducting tunnel junction} {\bf a}, Optical microscope image of one the samples colored for clarity. It consists of an EuS layer (blue, $13$~nm),  an Al strip (gray, $16$~nm) covered by AlO$_x$ ($\sim 4$~nm), and a cross bar of Co (red, $14$~nm). 
The circuit shows the 4 probe measurement setup utilized in our study, and the external magnetic field direction which is parallel to the Co strip.
{\bf b}, Schematic of the sample side view (top) and simplified representation with arbitrary scales of device density of states (bottom), with the spin-split Al superconducting gap (left) and spin-polarized Co 3d bands (right).
{\bf c}, Contour plot of the tunneling conductance ($G(V)=dI/dV$) measured vs the external magnetic field ($B$) and the voltage drop across the junction ($V_{\rm Co}-V_{\rm Al})$. The sweep direction of the magnetic field is indicated by the arrow. 
{\bf d}, Tunneling spectroscopy for parallel (P, green) and antiparallel (AP, orange) configurations at representative magnetic field strengths. The curves are extracted from (c) and indicated there with horizontal dashed lines in the respective colors. Continuous lines are the fit to the experimental data based on the tunneling model as described in the Methods section~\ref{Theory}.
{\bf e}, Tunneling magnetoresistance TMR=($R_{AP}$ - $R_{P}$ )/$R_{P}$, where $R_{AP}$ and $R_{P}$ are the junction resistances in the antiparallel and parallel configurations, respectively, extracted from (d) vs voltage drop. The red line is the theoretical expectation for the fits shown in panel (d).
{\bf f} Tunneling conductance $G(V)$ vs  external magnetic field. The vertical black (purple) dashed line in (c) indicates the positive (negative) excitation used to extract the magnetoconductance. Here, the magnetic field sweep directions are indicated by the arrows and a hysteresis is discernible, presenting a clear TMR behavior. All measurement were performed at a bath temperature $T_{bath}=100$~mK.
}
\label{fig:electric}
\end{figure*}

A promising alternative based on superconductor materials can represent a step forward in thermoelectric-based technology.
Differing from semiconducting materials, metals do not suffer charge freeze out, and semiconductor-like properties are still present in the quasi-particle spectrum characterized by the superconducting gap. 
Still, electron-hole asymmetry is difficult to achieve in conventional superconductors due to charge neutrality constraints, unless strong non-equilibrium conditions are present \cite{germanese2022bipolar,germanese2023phase}.
Recently, a superconducting spin-caloritronic scheme based on spin-selective tunnel junctions was proposed, enabling breaking the electron-hole symmetry while keeping charge neutrality and resulting in the generation of large thermoelectric effects~\cite{ozaeta2014predicted}. 
This prediction was confirmed by thermocurrents measured in superconducting tunnel junctions, with spin-split superconductors obtained via external fields \cite{kolenda2016observation}, and by exchange interactions~\cite{kolenda2017thermoelectric} present in thin superconductor/ferromagnetic-insulator (S/FI) bilayers~\cite{meservey_spin-polarized_1994,strambini_revealing_2017,hijano_coexistence_2021,de_simoni_toward_2018}. So far, no thermovoltage or resulting thermopower has been demonstrated despite its key role for energy harvesting at cryogenic temperatures and resulting applications for radiation detection\cite{heikkila2018thermoelectric,geng_superconductor-ferromagnet_2023}

Here, we implement a superconducting spin-selective tunnel junction based on a multilayer of EuS/Al/AlO$_x$/Co. A strong thermovoltage ($ \sim 10$ $\mu$V) is generated at sub-Kelvin temperatures ($<1$ K) with a magnitude close to its upper bound dictated by the Al superconducting gap ($\Delta \simeq 200$ $\mu$eV). The resulting Seebeck coefficient is of the order of few hundred of $\mu$V/K for different temperature and magnetic configurations. A sizable work was extracted by the junction therefore demonstrating a superconducting spintronic heat engine. Yet, the efficiency and functionality of the engine are quantified for different magnetic configurations. Fianlly, the implementation of a two-state thermoelectric memory cell based on the device magnetic hysteresis is discussed.

\section{Sample design and non-reciprocity}

The device consists of a superconducting thin film (aluminum-Al) proximitized by a ferromagnetic insulator (europium sulfide-EuS) on one side and separated from a ferromagnet (cobalt-Co) on the other side by an insulating barrier (aluminum oxide-AlO$_x$).

Figure \ref{fig:electric}a presents a micrograph of a typical device. A schematic of the four-wire measurement used for the tunneling spectroscopy is likewise shown.
In the cartoon of Fig.~\ref{fig:electric}b the side view of the sample is shown with a simplified representation of its density of states (DOS) on the bottom. 
The tunneling conductance of the device is strongly influenced by the exchange spin-splitting of the superconductor DOS (on the left) facing the spin-split DOS of the Co counter-electrode such that strong spin filtering is expected for proper voltage bias. 
Such spin filtering is at the origin of the asymmetric tunneling conductance $G(V)=dI/dV$ measured as a function of the bias voltage and magnetic field, and presented in the color plot of Fig.~\ref{fig:electric}c. 
The characteristic asymmetry in $G(V)$ can be seen in Fig.~\ref{fig:electric}d for the green line.
It is compatible with a parallel (P) alignment of the magnetizations of the EuS and Co layers, visible for most of the magnetic fields explored. Only the tunneling conductance at $B\simeq 10 $ mT, obtained during the sequential switching of the two ferromagnets, is characterized by an anti-parallel (AP) alignment (Fig.~\ref{fig:electric}c and d orange line). 
By fitting the experimental data with the spin selective tunneling model \cite{heikkila2019thermal} (see Methods section \ref{Theory} for model details and 
continuous lines in Fig.~\ref{fig:electric}d for the fit) it is possible to extract the spin polarization of the tunnel barrier $P\simeq0.5$, the exchange interaction induced in the Al layer $h\simeq50$ $\mu$eV, the Al superconducting gap $\Delta \simeq 195$ $\mu $eV, and the inelastic and spin-flip scattering rates $\hbar\Gamma\simeq 32$ $\mu $eV and  $\hbar\Gamma_{\rm sf}\simeq 29$ $\mu $eV, for $B=-35$~mT, respectively. 
Notably, a large $\Gamma$ characterizes the superconducting tunnel barrier, as typically observed in junctions with ferromagnetic counterelectrodes~\cite{li2013observation,munzenberg_superconductor-ferromagnet_2004}.
Different devices were tested showing similar results with slightly different $\Gamma$, $\Gamma_{\rm sf}$, $h$ and $P$ (see the extended Figure~\ref{fig:extfig2}).  
From the tunneling spectroscopy it is also possible to extract the field evolution of the tunneling magnetoresistance $TMR= \frac{R_P-R_{AP}}{R_{AP}}$, shown in Fig.~\ref{fig:electric}e, and the magnetoconductance, shown in Fig.~\ref{fig:electric}f.
Values obtained are compatible with previous $TMR$ measured in similar structures \cite{li2013observation}, showing a maximum at voltage biases compatible with the superconducting gap ($e V_{MAX} \simeq \pm \Delta$).
\begin{figure*}[hbt!]
\center
\includegraphics[width=0.95\textwidth]{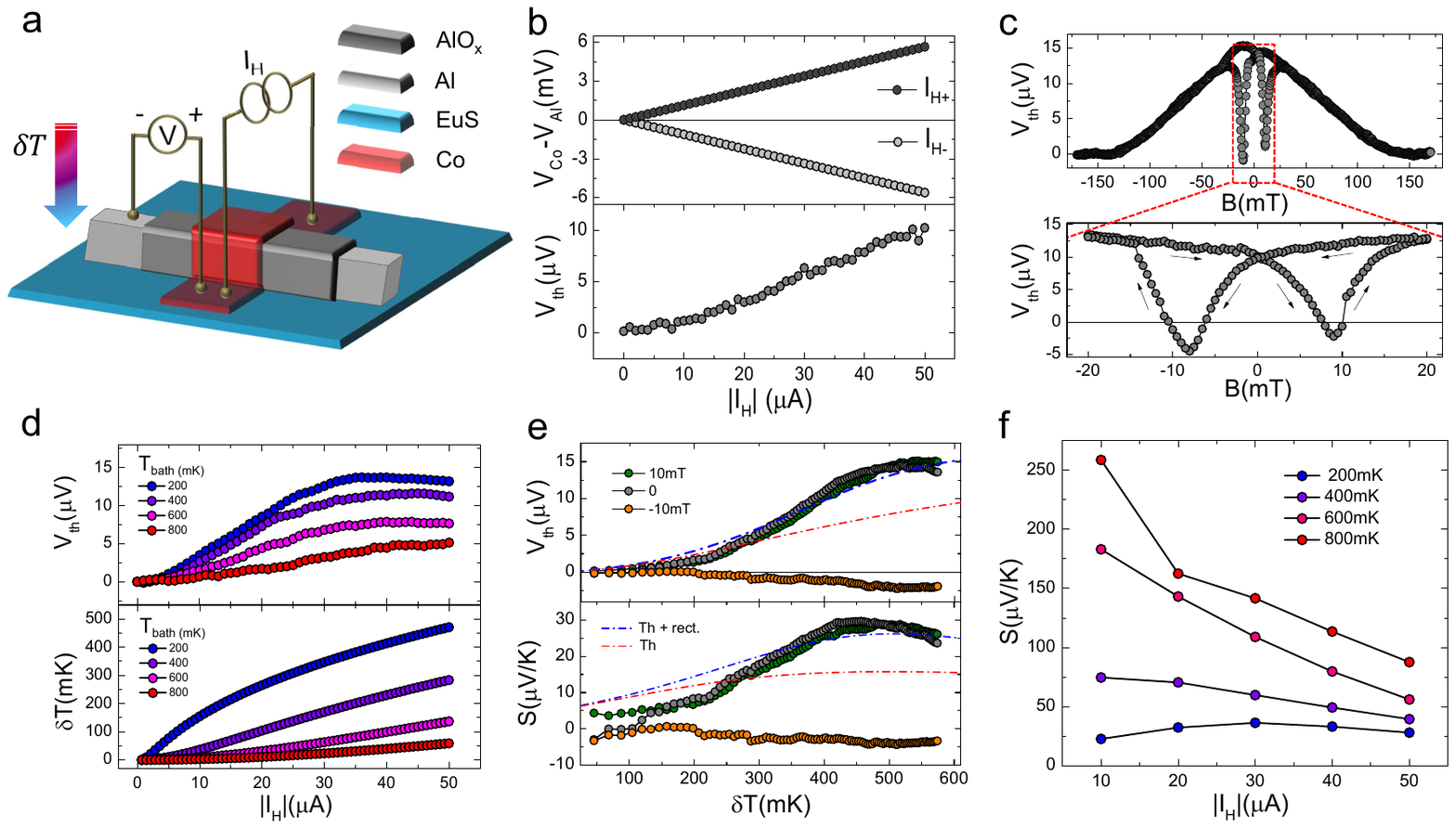}
\caption{{\bf Thermoelectric characterization.}
{\bf a},  Scheme of the electric circuit used to quantify the thermoelectric response of the device, with the temperature difference obtained via a Joule heating current flowing through the Co strip. 
{\bf b}, Voltage measured vs heating current, with the positive voltmeter electrode on Co and negative on Al at B=$10$mT . The thermovoltage is obtained by subtracting the Ohmic contribution using the average $V_{\rm th}= \frac{V(+I_{H})+V_(-I_{H})}{2}$. 
{\bf c}, $V_{\rm th}$ measured at $|I_H|=40$$\mu$A vs external magnetic field. 
{\bf d}, $V_{\rm th}$  measured vs $I_H$ for different bath temperatures up to $T=800$ mK (top panel) and temperature difference $\delta T = T_{Co}-T_{Al}$ estimated from model fits to the tunneling spectroscopy, performed at different bath temperatures and heating currents (bottom panel). 
{\bf e}, Thermovoltage (top) and Seebeck coefficient ($S$, bottom) vs temperature difference at selected magnetic field values measured at $T_{\rm bath}=100$mK. Dash-dotted lines represent the fits to the data considering thermoelectricity with (blue) and without (red) rectification effects.
{\bf f}, Seebeck coefficient $S$ extracted at different $T_{\rm bath}$ and $I_H$ 
}
\label{fig:thermovoltage}
\end{figure*}

\section{Thermoelectric response}

To quantify the thermoelectric response of the device, the thermovoltage across the junction was measured in the presence of a thermal gradient imposed across the junction. Such temperature difference is achieved via a Joule-heating current $I_H$ which flows through the Co strip while the Al is thermalized by the substrate at bath temperature ($T_{\rm bath}$), according to the scheme presented in Figure~\ref{fig:thermovoltage}a. 
The voltage measured across the junction was then symmetrized with respect to $I_H$ ($V_{\rm th}= \frac{V(+I_{H})+V(-I_{H})}{2}$) to remove the trivial ohmic contribution originating from the shared electrical paths between the voltage probe and the heating current, similarly to previous experiments on transversal rectification in superconducting tunnel diodes~\cite{strambini2022superconducting}. 
Differing from tunnel diodes, the larger impedance of the device makes thermoelectricity the main contribution of the voltage drop summing to rectification components.
A representative example of symmetrization is presented in Figure~\ref{fig:thermovoltage}b. In the top panel the voltage measured as a function of $|I_H|$ presents the main linear evolution, while only after symmetrization (bottom panel), small deviations are visible and a clear monotonic increase of $V_{\rm th}(I_H)$ up to $\simeq 10 \mu$V is observed at $I_H=50\mu$A.
Above 50$\mu$A the large power injected in the device is not fully dissipated by the substrate limiting the thermalization of the cold Al lead and resulting in a saturation or decrease of $V_{\rm th}$.
The increase of the Al temperature at large $I_H$ was confirmed by the damping of the critical current measured in the Al strip at different $I_H$ as shown in the extended figure~\ref{fig:extfig3}. 
The evolution of $V_{\rm th}(B)$ in the external magnetic field at fixed $|I_H|$ is shown in Fig.~\ref{fig:thermovoltage}c. 
Consistently with the non-reciprocal tunneling spectroscopy measurements, $V_{\rm th}$ strongly depends on $B$ and on the relative orientation of the two ferromagnetic layers showing sign reversal in the AP phase.
Hysteresis in the magnetic field is visible, with a maximum signal at $|B|\simeq 20$~mT vanishing above 120~mT due to the quenching of superconductivity, as observed also in the tunneling spectroscopy measurements reported in fig.~\ref{fig:electric}c. 
In the inset, showing the central measurement range, it is possible to appreciate the sizable signal ($>10~ \mu$V) present even at zero field as a consequence of the strong ferromagnetism of the device. Moreover, a clear negative thermovoltage is visible between the coercive fields of the two ferromagnetic layers ($7$ mT$\lesssim |B|\lesssim 10$ mT). 
Such inversion of $V_{\rm th}$ confirms the AP phase achieved between the Co and EuS ferromagnetic layers as deduced from the tunneling spectroscopy in the same field range.
The lower amplitude of $V_{\rm th}$ in the AP phase with respect to the P case is consistent with a weaker polarization and spin filtering of the device, and it indicates a partially polarized magnetization of the EuS and Co layers in the AP phase during the non simultaneous magnetization switching of the two ferromagnets.
At higher temperatures $V_{\rm th}$ tends to slowly decrease as shown in the top panel of Fig.~\ref{fig:thermovoltage}d with a sizable thermovoltage observed up to 800 mK.
Such robustness in temperature is a consequence of the large exchange splitting of the device $h\simeq k_B\times{}600$mK extending the operation of the device to higher temperature, see Extended Data Fig.~\ref{fig:extfig1}. 
To evaluate the Seebeck coefficient from $V_{\rm th}$ the temperature gradient across the junction $\delta T = T_{\mathrm{Co}} - T_{\mathrm{Al}}$ needs to be estimated. The thermal model for the device (see Methods) 
indicates that at low heating power $T_{\mathrm{Al}}\simeq T_{\mathrm{bath}}$, as confirmed also by monitoring the critical current of the Al lead at different $I_H$ (see extended figure~\ref{fig:extfig3}). 
The temperature of the Co electrode is estimated from the broadening of the tunneling spectroscopy as typically done in S/I/N thermometry \cite{giazotto_opportunities_2006} and was measured at different $I_H$. In this case, the model needs to be extended to account also for the additional lateral voltage drop due to the presence of the heating current $I_H\ne0$, (see Methods for details). The full model provides the estimate $\delta T(I_H,T_{\mathrm{bath}}) = (T_{\mathrm{bath}}^5 + b I_H^2)^{1/5} - T_{\mathrm{bath}}$ with $b\approx5.6\cdot10^{-5}\,\mathrm{K^5/\mu{}A^2}$, shown in the bottom panel of Fig.~\ref{fig:thermovoltage}d. Notably, a large temperature gradient up to $500$~mK was achieved across the junction for low $T_{\mathrm{bath}}$.
Using the relation $\delta T (I_H,T_{\mathrm{bath}})$ it is possible to remap the thermovoltage in the temperature gradient $V_{\rm th}(\delta T)$ as shown in Fig.~\ref{fig:thermovoltage}e together with the resulting Seebeck coefficient $S=V_{\rm th}/\delta T$. A Seebeck coefficient of up to $30\mu$V/K can be estimated both at 10 mT and at zero magnetic field for a base temperature of $100$mK, which is on par with state-of-the-art cryogenic thermoelectric elements~\cite{kolenda2016observation,germanese2022bipolar,germanese2023phase,josefsson_quantum-dot_2018}. It is worthwhile to mention that by increasing the bath temperature above $500$mK, $S$ can obtain values as large as a few hundreds of $\mu$V/K.  
Moreover, a smaller but sizable Seebeck coefficient of $-5~\mu$V/K is also visible in the AP phase obtained at $B=-10$mT, thus implementing a thermoelectric spin valve where the \textit{n}-type and \textit{p}-type Seebeck effect is controlled by the relative orientation of the device magnetic moments.

\begin{figure*}[hbt!]
\center
\includegraphics[width=0.95\textwidth]{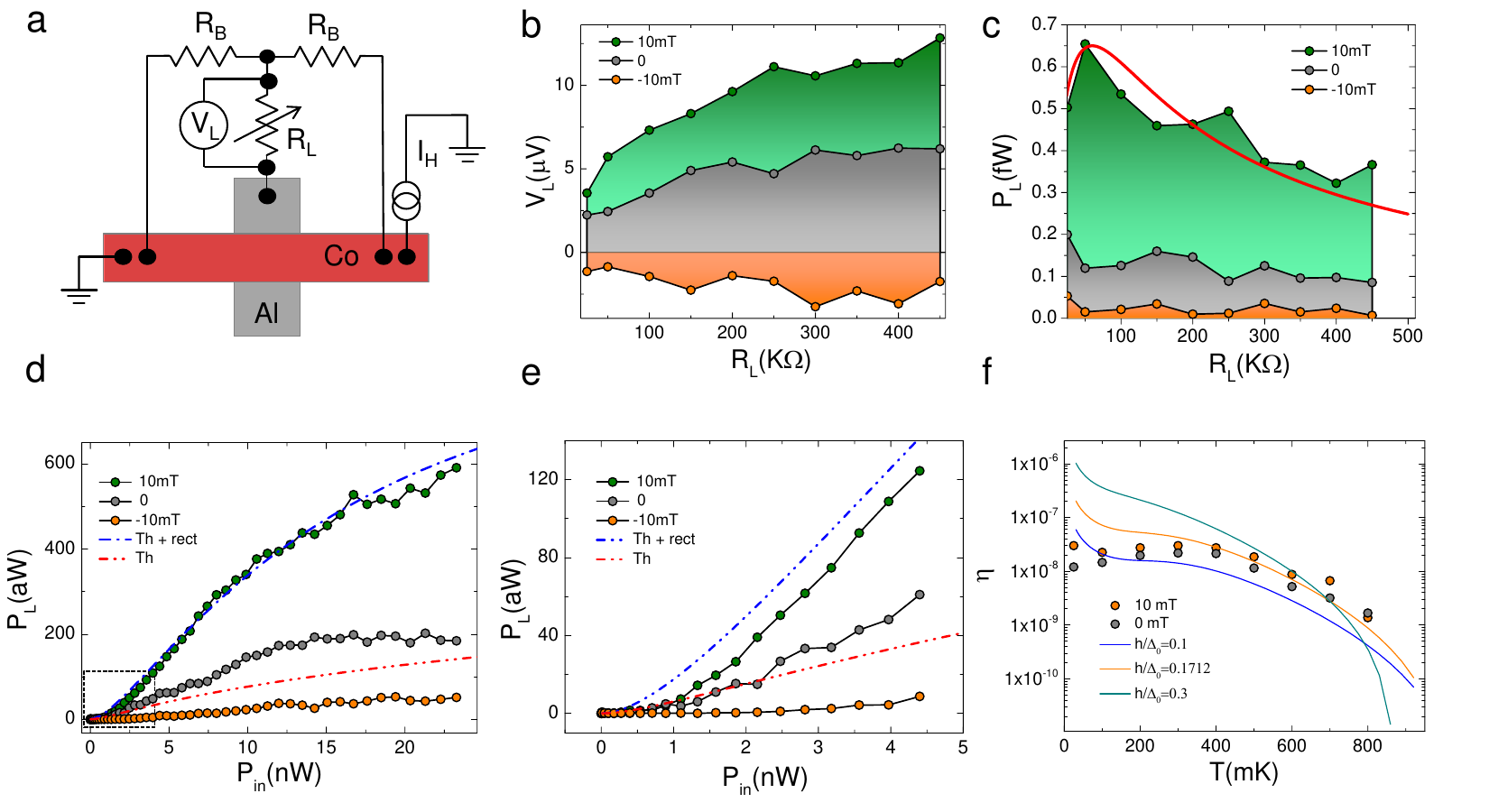}
\caption{{\bf Heat engine characterization.}  
{\bf a} Circuit utilized for the heat engine measurement with the heating current ($I_H$) flowing through the cobalt strip.
The tunnel junction is shunted with a variable load resistor ($R_L$) and two symmetric balancing resistances  ($R_B\ll R_L$) to avoid spurious leaks of  $I_H$ in the load.
{\bf b} Voltage developed across the load ($V_L$) measured at different $R_L$ for $|I_H|=40~\mu $A at $T_{\rm bath}=25$ mK and three  magnetic field values corresponding to different regimes, i.e., P saturation at  10~mT, remanence at 0~mT, and AP configuration at -10~mT.
{\bf c} Power load ($P_L=\frac {V_L^2}{R_L}$) extracted from panel (b) vs  $R_L$. The red curve is a fit to the 10~mT data considering the maximum transferred power satisfying $P_L = I^2 R_L = \frac{R_LV_{S}^2}{(R_L+R_S)^2}$ for a voltage source of $V_S=V_{\rm th}\simeq 12.5~\mu$V, and an internal resistance $R_S \simeq 60$ k$\Omega$. 
{\bf d} Evolution of $P_L$ measured at  $R_L=150$ k$\Omega$ and $T_{\rm bath}= 25$ mK vs input power  ($P_{\rm in}=I_H^2/R_{\rm Co}$), with $R_{\rm Co}=11$ $\Omega$ corresponding to the resistance of the Co strip on top of the tunnel junction.
{\bf e} Blow-up of the $P_L$  behavior of panel (d) in a restricted range of $P_{\rm in}$. Dash-dotted lines represent the fits to the data considering thermoelectricity with (blue) and without (red) rectification effects.
{\bf f} Efficiency $\eta = \frac {P_{L}}{P_{\rm in}}$, measured as a function of temperature $T$ at $R_L=150$ k$\Omega$.
 Full lines are the theoretical prediction for $\eta$ in the linear-response approximation (see Methods section \ref{Theory} for details), and evaluated for the junction parameters used in the fit of Fig.~\ref{fig:electric}d.
}
\label{fig:schematics}
\end{figure*}
\begin{figure*}[hbt!]
\center
\includegraphics[width=0.95\textwidth]{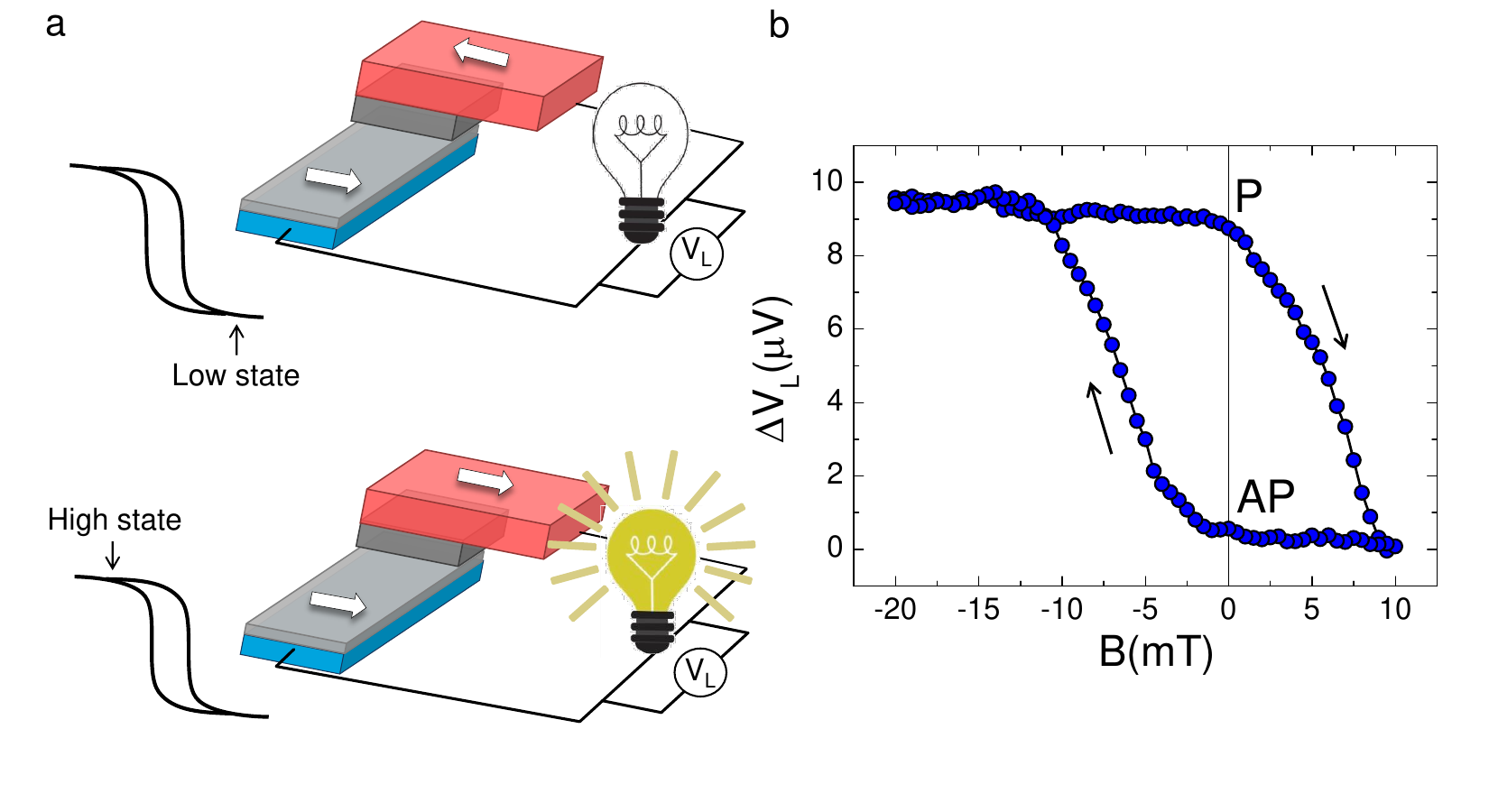}
\caption{{\bf Thermoelectric memory cell.} 
{\bf a} Simplified scheme representing the hysteretic behavior of the thermoelectric memory cell investigated with a low state represented by the AP configuration between the spin-split superconductor and the ferromagnetic cobalt layer, and a high state when the magnetization on the device electrodes are P and a larger voltage is generated across the load resistor. 
{\bf b} Evolution of the differential voltage drop across the 
load ($\Delta V_L=V_L^{P}-V_L^{AP}$) measured in the memory cell vs magnetic field $B$ with a load resistor $R_L=150$ k$\Omega$ at $T_{\rm bath}=300$ mK. Note the sizable signal contrast $\sim 10\mu$V the memory cell can provide even at zero magnetic field. 
}
\label{fig:memory}
\end{figure*}

\section{Heat Engine}

Once $V_{\rm th}$ is applied on a load resistor $R_L$, work can be extracted from the thermoelectric effect for the demonstration of a heat engine, and we can quantify the thermal-to-electrical energy conversion.
The circuit used for this purpose is sketched in Fig.~\ref{fig:schematics}a. The junction is shunted to the ground with $R_L$ and two additional balancing resistors ($R_B = 10~k\Omega$) have been included in the circuit in a symmetric configuration to prevent spurious leaks from the heating-current source to the load.
The voltage drop ($V_L$) across $R_L$ is probed for different measurement configurations and the resulting dissipated power ($P_L= \frac {V^2_L}{R_L} $) is used to estimate the power generated by the engine, then neglecting residual thermoelectric power dissipated in the balancing resistors $R_B$.   
The evolution of $V_L$ measured as a function of $R_L$, under constant heating with $|I_H|=40~\mu $A is presented in Fig.~\ref{fig:schematics}b.
Three different magnetic configurations are compared, P saturation regime (green, $B=10$~mT), P remanence (grey, $B=0$~mT) and AP (orange, $B=-10$~mT) characterized by a negative $V_L$ as for the thermovoltage shown in Fig.~\ref{fig:electric}c.
In all presented field regimes $V_L$ tends to increase with an increasing $R_L$, showing a saturation towards the thermovoltage $V_{\rm th}$ when the load resistance is above 500~k$\Omega$, i.e., much larger than the tunnel resistance ($R_L \gg R_T\simeq 50$~k$\Omega$) as expected for an ideal voltage source $V_L= V_{\rm th} \frac{R_L}{R_L+R_T}$, with $R_T$ as source resistance. 
The resulting dissipated power shown in Fig.~\ref{fig:schematics}c has a different non-monotonic behavior with a maximum observed for $R_L \sim R_T$.
This behavior is consistent with the maximum power transfer theorem (known as Jacobi's law) predicting $P_L = I^2 R_L = \frac{V_{L}^2 R_L}{(R_L+R_T)^2}$ 
as reported in the red fit line in Fig.~\ref{fig:schematics}c.

The heat engine efficiency can be quantified from the ratio between the power extracted by the engine ($P_L$) and the Joule power injected to generate the thermal gradient $P_{\rm in} = R_{\rm Co} \times I_H^2$, where $R_{\rm Co}=11$ $\Omega$ 
is the resistance of the cobalt strip at the junction. 
Figures \ref{fig:schematics}d and \ref{fig:schematics}e show $P_L$ vs $P_{\rm in}$ for the three magnetic field configurations. 
At low power $P_L$ is characterized by an almost linear increase corresponding to an efficiency $\eta = P_L/P_{\rm in} \simeq 5 \times 10^{-8}$ that tends to decrease at high power.
We stress that this efficiency is just a lower bound of the intrinsic efficiency of the effect, as most of $P_{\rm in}$ may be lost in different heat channels including the heat dispersed from the Co directly to the substrate. 
The temperature dependence of $\eta$ displayed in Fig.~\ref{fig:schematics}f shows an almost constant efficiency below 400~mK, and a quick damping at higher temperatures.
The quick damping is consistent with the expected behavior of $\eta(T)$ obtained from the theoretical model shown in Fig.~\ref{fig:schematics}f based on the device parameters, as described in detail in the methods section. 
The model shows that the main limiting factor for $\eta$ at high temperatures is the electron-phonon coupling: most of the thermal energy from the electrons is transferred to the lattice phonons, instead of being converted into thermoelectric power. 
In fact, the thermal conductance between electrons and lattice phonons scales as $\propto{}T^4$ \cite{giazotto_opportunities_2006,fornieri2017towards}, leading to a decreasing efficiency at increasing temperature.
Yet, as observed from the theoretical model, the heat engine efficiency could be strongly enhanced by working below 100 mK.

\section{Heat-engine memory}
It is worthwhile noting that combining the hysteretic behavior of the thermoelectric effect as shown in Fig.~\ref{fig:electric}f with the heat engine, it is possible to envision a thermoelectric memory cell.
Our device structure is an original concept for a  classical memory cell similar to a conventional magnetic random access memory stack \cite{bhatti2017spintronics}. The latter, is composed by two ferromagnetic layers with different coercivity, and separated by a tunnel junction. The first ferromagnetic electrode provides the electronic spin polarization in two Mott channels \cite{mott1936electrical}, while the second feromagnetic layer filters the spin polarized currents after coherent tunneling, thereby allowing high conductance in P configuration and lower conductance in the AP state.
By contrast, in our  memory cell the logic states are codified by a thermoelectric voltage self generated in the memory itself, and not requiring a local input current for the state read-out. This advantage can strongly simplify the wiring of the memory with net benefits for scalability. 
Additionally, it allows for a local direct transduction of the electrical signal in to another physical observable, for instance, a photon in the sketch of Fig.~\ref{fig:memory}a. This may inspire novel methods for read-out and packaging in dense arrays. 
In Fig.~\ref{fig:memory}b, we present an example of the heat-engine hysteretic cycle measured in the proposed memory cell.
The device is first polarized in the P state at -20~mT, then $B$ is cycled between $-20$ mT to 10~mT. 
A clear hysteretic loop is visible with a high contrast of 10 $\mu$V between the P and AP configuration also at zero field, an important condition to operate the memory in the absence of external magnetic fields.

\section{Discussion and Conclusions}\label{sec12}

In summary, we have fabricated and 
 characterized a superconductor-ferromagnet tunnel junction structure based on aluminum proximitized by europium sulfide, and separated from a cobalt electrode by an aluminum oxide tunnel barrier. 
Our device shows a remarkable non-reciprocal charge transport due to electron-hole symmetry breaking induced by the spin selectivity of the junction. 
As a consequence, a sizable thermoelectric voltage is observed in the presence of a thermal gradient, which is achieved via a Joule-heating current flowing through  the cobalt strip. The different coercivity of the two ferromagnetic layers exploited in the junction joined to the large ferromagnetic remanence of Co magnetization warrant two important features: (i) thermoelectricity is observed even at zero magnetic field and (ii) a clear inversion of the thermoelectric effect is achieved when the EuS and Co magnetizations are antiparallel. The latter implements the first spin valve for thermoelectric applications, by reversing the Seebeck coefficient from \textit{p}-type to \textit{n}-type. 
We quantified the power generated by the structure  over a series of external load resistors thereby demonstrating the implementation of a superconducting spintronic heat engine. 
From the thermal model of the device we identified the heat losses through the electron-phonon coupling as the main factor limiting the engine efficiency.

In light of future technological applications, several strategies could be followed in order to increase the engine efficiency, such as decreasing the junction size, and improving  heat isolation via device suspension or by lowering the operating temperature in order to limit the electron-phonon coupling.
Finally, we have also shown the operation of
the device as a thermoelectric superconducting memory cell. For that purpose, two main advantages are envisioned: (i) junction durability, if operated in an open circuit configuration with no detrimental currents flowing through the junction; (ii) high scalability, due to a read-out signal self generated by the thermoelectric power.
We envision the application of such cryogenic thermoelectric element in the implementation of sensitive self-biased detectors of electromagnetic radiation with simplified approaches for multiplexing~\cite{geng_superconductor-ferromagnet_2023,heikkila_thermoelectric_2018,ilic_apparatus_2023}. 
Additionally, by scaling our heat engine to large areas may find relevant applications for energy harvesting in the deep space where the low temperature makes conventional approaches somewhat ineffective.

\section{Methods}
\subsection{Sample preparation and measurement}
\label{Sample perpar}
For the sample preparation we have first deposited $12.5$ nm EuS thin film by molecular beam epitaxy on top of Si/SiO$_x$ substrates cooled at $150$K. The pressure during growth was kept in the range of $10^{-9}$~mbar to avoid any EuS oxidation and achieve a near stoichiometric EuS compound. 
Without breaking the ultra high vacuum conditions, a $\sim250-\mu$m-width Al lead was grown on top by using a metallic shadow mask. 
The total thickness of the Al layer was $20$~nm. To form the insulating AlO$_x$ barrier the sample was exposed $3 \times 10^{-3}$~mbar of low-energy oxygen plasma created by inductevely coupled plasma source for 5 hours, resulting in a $\sim 4$-nm-thick AlO$_x$ layer and lowering the metallic Al thickness down to $16$~nm.
The subsequent cross bar geometry was realized with another shadow mask evaporation to grow a Co lead of 14~nm thickness and $\sim 200~\mu$m width. 
Al and Co layers were grown with e-beam metal evaporators. Finally, before extracting the sample from the chamber, a $7$~nm calcium fluoride (CaF) layer was deposited covering the whole system to avoid environmental oxidation. 
Samples were wire bonded with aluminum wires and mounted in a dilution fridge, where the magneto-electrical measurements were performed through low pass filters. All the signals were amplified via low-noise voltage and current preamplifiers. 
For the application of the heating currents  and critical current measurements the filters were bypassed in order to decrease the power load that would provide cryostat excessive heating. 

\subsection{Theoretical model}
\label{Theory}

The current density through the Al/Co tunnel junction can be expressed as \cite{ozaeta2014predicted,heikkila2019thermal},
\begin{align}
  \mathcal{I}
  &=
  \sum_{\sigma=\uparrow,\downarrow}
  G_\sigma^\square
  \int_{-\infty}^\infty dE\,
  N_{\sigma}(E)[f(E,T_S) - f(E + V,T_N)]
  \,,
  \label{eq:IV-theor}
\end{align}
and depends on the voltage $V$ over the junction and temperatures $T_N$, $T_S$ on the normal (Co) and superconductor (Al) sides.
Here $f(E,T)$ is a Fermi function. The current is proportional to spin-dependent conductances per square area $G^\square_{\uparrow/\downarrow}=\frac{1\pm P}{2}G_\square$,
which are due to Co spin polarization and interface properties. Here, $-1\le{}P\le1$ is the spin polarization and $G_\square$ the junction conductance per square.
The result also depends on the superconductor density of states $N_\sigma(E)$. The superconducting gap $\Delta$ in it is spin-split by an exchange field $h$ induced from EuS, but this is counteracted by spin-flip scattering with rate $\Gamma_{\rm sf}$ and inelastic scattering with rate $\Gamma$, which we account for using methods in Ref.~\cite{heikkila2019thermal}. We extract the values of $\Delta$, $\Gamma_{\rm sf}$, $\Gamma$, $G$, $P$, and $h$ by fitting Eq.~\eqref{eq:IV-theor} to experimental $I(V)$ characteristics at $T_S=T_N$.

As the Al/Co tunnel junction resistance is high compared to the total Co wire resistance, the voltage profile along the Co wire is linear to a good approximation, $V(x)=V_0 + x I_H R_x/L_x$, where $R_x=\rho_{\mathrm{Co}} L_x /(W t_{\mathrm{Co}})$ is the lateral resistance of the part of the Co film (cross-section $t_{\mathrm{Co}}\times{}W$, length $L_x$, resistivity $\rho_{\mathrm{Co}}$) on top of the tunnel junction. Due to superflow the voltage in the Al film is spatially constant. The total tunneling current through the Al/Co junction then is

\begin{align}
 \label{eq:I-tunnel}
 I_T = W\int_{-L_x/2}^{L_x/2}dx\,\mathcal{I}(V(x),T_{\mathrm{Co}},T_{\mathrm{Al}})
\end{align}
where the Al/Co overlap has size $W\times{}L_x$ and $\mathcal{I}$ is the local S/I/FM junction current density--voltage relation from Eq.~\eqref{eq:IV-theor}, which includes the S/FI thermoelectric effects. \cite{ozaeta2014predicted} Under these conditions, the Joule heating via current $I_H$ in Co at low temperatures is mainly limited by the electron-phonon coupling. The corresponding heat balance equation is, \cite{giazotto_opportunities_2006}
\begin{align}
 \label{eq:heat-balance}
 L_xWt_{\mathrm{Co}} \Sigma (T_{\mathrm{Co}}^5 - T_{\mathrm{bath}}^5)
 = R_x I_H^2
 \,,
\end{align}
where $\Sigma$ the electron-phonon coupling parameter of Co. We have also modeled the Al side with a similar equation, with tunneling current input power on right-hand side and taking superconductivity into account. \cite{giazotto_opportunities_2006,ozaeta2014predicted,heikkila2018thermoelectric} We find that due to the high tunnel resistance, electronic heat transport across the tunnel junction is suppressed, and we can neglect heating of the Al side.
Consequently $\delta T = T_{\mathrm{Co}} - T_{\mathrm{Al}} = (T_{\mathrm{bath}}^5 + b I_H^2)^{1/5} - T_{\mathrm{bath}}$, with $b=R_x/(W t_{\mathrm{Co}}L_x\Sigma)$.
To determine the effective values of $\Sigma$ and $R_x$, a two-parameter fit of Eqs.~\eqref{eq:I-tunnel},\eqref{eq:heat-balance} is done on the experimental $dI/dV$ curves of different $T_{\mathrm{bath}}$ and $I_H$.
In addition, fits at $I_H=0$ are used to determine the tunnel junction parameters in Eq.~\eqref{eq:IV-theor}.

The circuit model of Fig.~3a together with Eqs.~\eqref{eq:IV-theor},\eqref{eq:I-tunnel},\eqref{eq:heat-balance} from the model of the heat engine, from which efficiencies and relative contributions of the rectification and thermoelectricity can be estimated. For small $I_H$, we can expand $I_T\approx G V_{\mathrm{th}} + P\alpha\delta T/T + \frac{G'}{24}(R_xI_H)^2$ where $G$ is the tunnel junction conductance at zero bias, $\alpha$ the thermoelectric coefficient, \cite{ozaeta2014predicted} and $G'$ the zero-bias voltage derivative of the conductance, characterizing the rectification.  In the tunneling model, $G'\approx{}ce^2P\alpha/(2k_B^2T^2)$ where $c\approx1$ is a weakly temperature-dependent numerical factor. From Eq.~\eqref{eq:heat-balance} one can then deduce that for high-resistance tunnel junctions, the thermoelectric contribution dominates when $T_{\mathrm{bath}}\lesssim[9k_B^2/(e^2\rho_{\mathrm{Co}}\Sigma L_x^2)]^{1/3}\approx300\,\mathrm{mK}$ and $I_H\lesssim25[k_B^5/(e^5R_x^4Wt_{\mathrm{Co}}\Sigma L_x)]^{1/3}\approx20\,\mathrm{\mu{}A}$. The opposite limit of low-resistance junctions was discussed in Ref.~\cite{strambini2022superconducting}.

The thermoelectric coefficient $\alpha$ is related to the Seebeck coefficient by $S=V_{\rm th}/\delta{}T=P\alpha/(GT)$,
and its temperature and exchange field dependence was discussed in Ref.~\cite{ozaeta2014predicted}.
The large value of $\Gamma$ in the experiment modifies the temperature and exchange field dependence,
as illustrated in the theoretical prediction in Extended Data Figure~\ref{fig:extfig1}.

\bibliography{reference}

\section*{Data availability} 
The data that support the findings of this study are available from corresponding author C.I.L.A., F.G and E.S. upon reasonable request. 

\section*{Code availability} 
The codes that support the findings of this study are available from corresponding author P.V. upon reasonable request.

\section*{Acknowledgments}

C.I.L.A., P.V., T.H. and F.G. acknowledge funding from the  EU's Horizon 2020 Research and Innovation Program under Grant Agreement No.~800923 (SuperTED). M.S. and E.S. acknowledge funding from the European Union’s Horizon 2020 research and innovation programme under the Marie Skłodowska Curie Action IF Grant No. 101022473 (SuperCONtacts).
F.G. and E.S. acknowledge the EU’s Horizon 2020
Research and Innovation Framework Program under
Grant Agreement No. 964398 (SUPERGATE), No.
101057977 (SPECTRUM), and the PNRR MUR
project PE0000023-NQSTI for partial financial support.
C.I.L.A. acknowledge Brazilian agencies FINEP, FAPEMIG APQ-04548-22, CNPq and CAPES (Finance Code 001).

\section*{Author contributions}
C.I.L.A., M.S. and E.S. performed the experiment and analysed the data. 
P.V. and T.T.H. provided theoretical support. 
C.G.O., S.K., M.I. and C.R. fabricated the samples. 
E.S. conceived the experiment together with F.G. and T.T.H..
C.I.L.A., M.S., P.V and E.S. wrote the manuscript with feedback from all authors. 

\section*{Competing interests}
The authors declare no competing interests.







\clearpage
\appendix
\section*{Extended figure}

\setcounter{figure}{0}
\makeatletter
\def\fnum@figure{{\bf Extended Data Figure \arabic{figure}:}}%
\def\@caption@fignum@sep{}%
\makeatother

\begin{figure*}
    \includegraphics{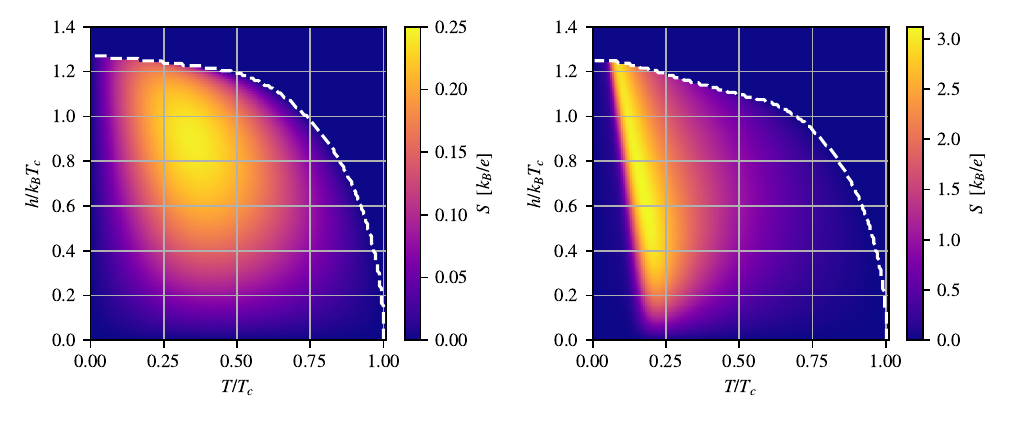}
    \caption{\label{fig:extfig1}
    {\bf Predicted Seebeck coefficient.}
    $S$ as a function of the exchange field $h$ and temperature $T$.
    Left: for the same parameters as in Fig.~2.
    Right: for $\Gamma_{\rm sf}=0$ and small $\hbar\Gamma=0.05\Delta$.
    Dashed line indicates the superconducting transition.
    }
\end{figure*}

\begin{figure*}
    \includegraphics[width=0.95\textwidth]{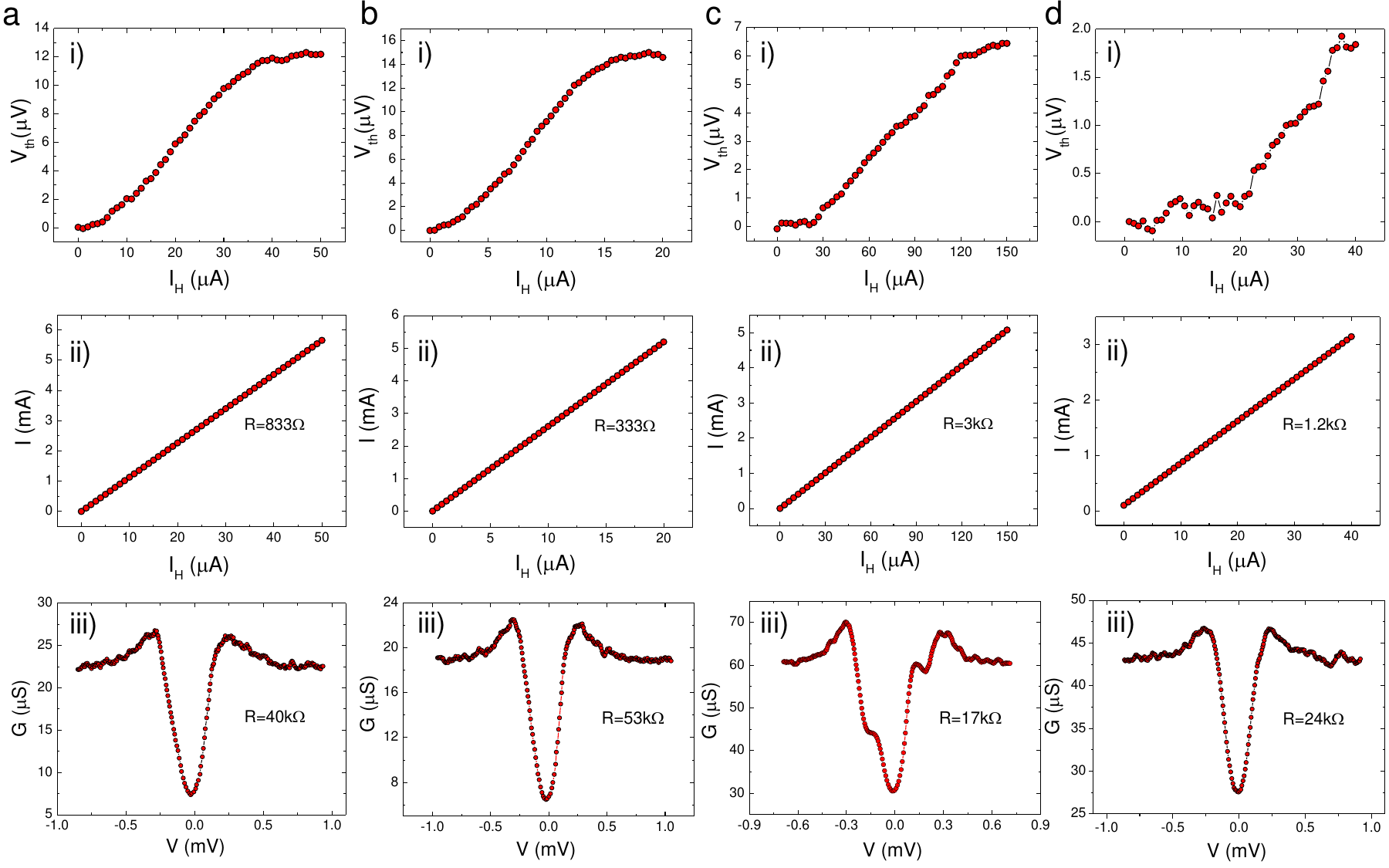}
    \caption{\label{fig:extfig2}
    {\bf Thermovoltage measurements on different samples.}
    We present in i) open circuit voltages in function of heating currents, performed at $T_{\rm bath}=100$~mK, for different samples. The cobalt strip resistances and junction resistances for the different devices are presented subsequently in ii) and iii). Panels a), b) and c) are related to different junction developed with aluminum thickness of 20nm, while in panel d) we present the results for a sample with aluminum thickness of 12nm.
    CaF/Co(14nm)/AlO$_x$ (5hrs)/Al(20 nm)/EuS(21.5nm)/SiO2/Si
    }
\end{figure*}

\begin{figure*}
    \includegraphics[width=0.95\textwidth]{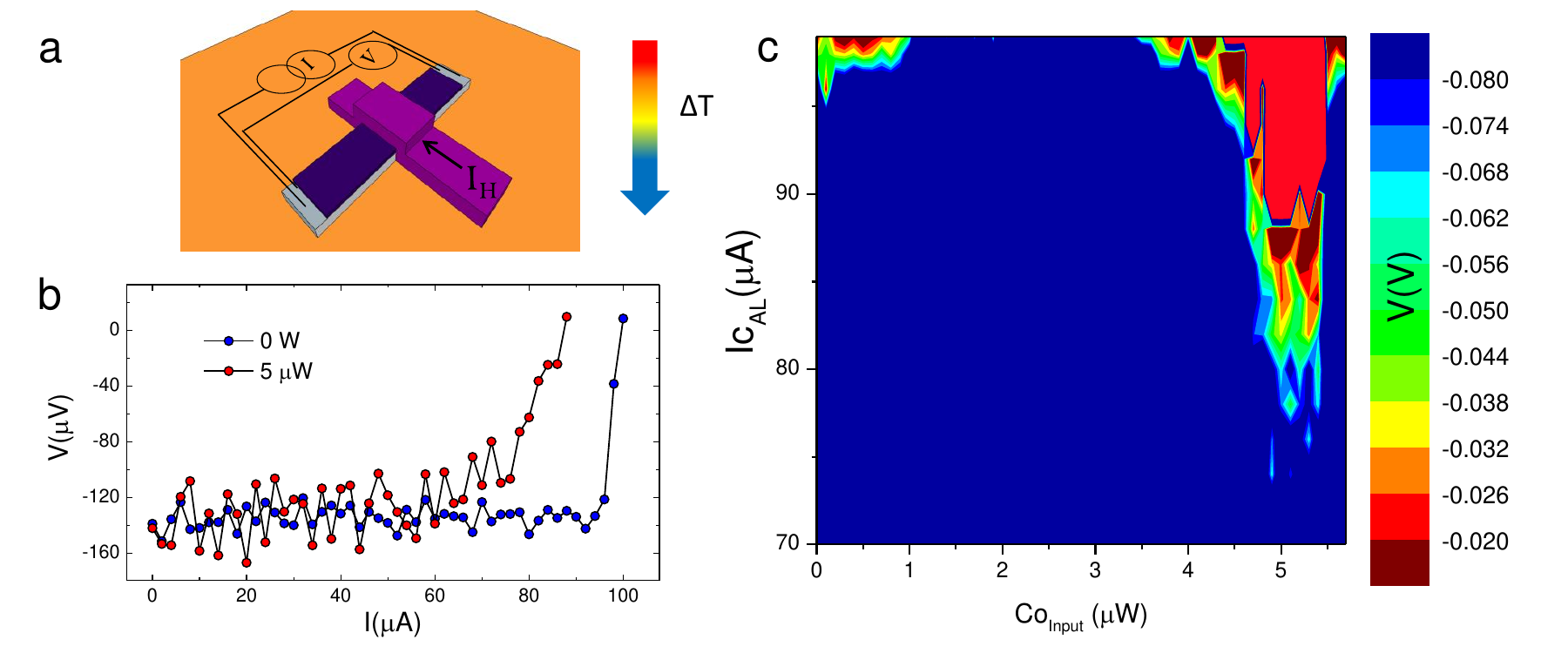}
    \caption{\label{fig:extfig3}
    {\bf Thermal gradient limit.}
    Critical current in the superconducting aluminum in function of power injected in the top cobalt strip, as depicted in the cartoon presented in a). b) examples of critical currents measured for the sample without excitation (blue dots) and under a power of $5 \mu W$, input in the top cobalt strip (red dots), the low negative voltage signal is due to a small offset in the voltage amplifier used. In panel c) we present a summary of critical current evolution in function of power, here it is possible to notice that the power generates a temperature gradient on the junction, but it does not affect the aluminum strip properties for the lower regime of power that we have used during the thermal effects presented in the paper.     
    }
\end{figure*}

\end{document}